\documentclass[prl,twocolumn,lengthcheck]{revtex4-1}

\usepackage{amsmath}
\usepackage{amssymb}
\usepackage{graphicx}
\usepackage{hyperref}
\usepackage{color}
\usepackage{braket}

\newcommand{\tr}{\operatorname{tr}}

\newcommand{\ic}{\ensuremath{i}}
\renewcommand{\d}{\ensuremath{d}}

\newcommand{\vk}{\ensuremath{\vec{k}}}

\begin{document}

\title{Entanglement renormalization for quantum fields}

\author{Jutho Haegeman$^{1}$}
\author{Tobias J.\ Osborne$^{2}$}
\author{Henri Verschelde$^{1}$}
\author{Frank Verstraete$^{3,4}$}
\affiliation{$^{1}$Ghent University, Department of Physics and Astronomy, Krijgslaan 281-S9, B-9000 Ghent, Belgium\\
$^{2}$Institut f\"ur Theoretische Physik, Leibniz Universit\"at Hannover, Appelstr. 2, 30167 Hannover, Germany\\
$^{3}$University of Vienna, Faculty of Physics, Boltzmanngasse 5, A-1090 Wien, Austria\\
$^{4}$C.N. Yang Institute for Theoretical Physics, SUNY, Stony Brook, NY 11794-3840, USA}

\begin{abstract}
It is shown how to construct renormalization group flows of quantum field theories in real space, as opposed to the usual Wilsonian approach in momentum space. This is achieved by generalizing the \emph{multiscale entanglement renormalization ansatz} to continuum theories. The variational class of wavefunctions arising from this RG flow are translation invariant and exhibit an entropy-area law. We illustrate the construction for a free non-relativistic boson model, and argue that the full power of the construction should emerge in the case of interacting theories.
\end{abstract}

\maketitle

Classical statistical mechanics, quantum many-body systems, and relativistic quantum field theories all involve an extremely large number of degrees of freedom living at different length scales. The interactions between these degrees of freedom are the source of notorious difficulties in their study. However, much insight has been gained from the renormalization group (RG), which has proven to be the natural tool to deal with the different length scales in such systems \cite{wilson}. In its original development, the RG acts as a fixed operation at the level of the classical partition function (or related quantities such as the effective action). This operation is typically formulated in momentum space and can only be implemented exactly for free theories. Perturbative expansions in a small parameter are required for interacting theories. In addition, this formulation of the RG is only applicable to quantum systems using the quantum-to-classical mapping, which is known to fail in some cases \cite{vojta}.

One exception is Wilson's \emph{numerical renormalization group} \cite{wilson}, which can be interpreted as an implementation of the RG directly at the level of the quantum wave function. Together with White's more powerful \emph{density matrix renormalization group} (DMRG) \cite{DMRG}, these methods are now understood as a variational optimization over the class of \emph{matrix product states} (MPS) \cite{MPS}. Based on the observation of an \emph{entropy/area law} \cite{arealaws} in \emph{locally} interacting quantum lattices, quantum-information-theoretical considerations have resulted in the development of other sophisticated variational classes for quantum lattice systems. These are generally known as tensor network states and can be associated with RG schemes, allowing the classification of gapped phases of matter \cite{tensornetworkrg}. Unlike Wilson's fixed RG scheme, these are variable RG schemes that can be variationally optimized. They are formulated in real-space and deal equally well with free and interacting systems. One specific scheme, called \emph{entanglement renormalization} \cite{ER}, can also be applied to study critical phases and can be used to compute, \text{e.g.}, scaling exponents \cite{merascaling}. The corresponding variational class, the \emph{multiscale entanglement renormalization ansatz} (MERA) \cite{mera}, is set apart by its unique properties, including, the ability to support algebraically decaying correlations and logarithmic corrections to the entropy/area law in $(1+1)$ dimensions. This class has been successfully applied to study the physics of a wide variety of strongly interacting systems in low dimensions, including the study of real-time evolution, and fermionic and anyonic systems which are inaccessible by Monte Carlo methods \cite{mera,merageneral,mera2d,merafermions}. 

Most of these developments have been restricted to the lattice setting. While they do allow the study of continuous quantum systems via discretisation, it is often desirable to work directly in the continuum. Examples in condensed matter physics include strongly interacting ultracold atomic gases \cite{Bloch} and impurity problems \cite{wilson, Impurities}, whereas the fermion doubling problem \cite{fermiondoubling} clearly motivates a continuum treatment of relativistic theories.  Recently it was understood how to define MPS and its higher-dimensional analogue for continuous systems \cite{cMPS}. This class has already been successfully applied to study both non-relativistic and relativistic quantum field theories \cite{RelcMPS}. It is the objective of this Letter to define a generalisation of MERA directly in the continuum. The area law for entanglement entropy ---with logarithmic violations for critical theories in $(1+1)$ dimensions--- has also been observed in the continuum in the context of black hole physics \cite{blackhole} and conformal field theories \cite{cft}, and thus validates the potential usefulness of our approach. 

\paragraph{The MERA class.---}\hspace{-1em} The MERA construction, introduced in \cite{mera}, may be described either via an \emph{active} renormalization process applied to a strongly correlated \emph{quantum state} or, \emph{dually}, as the result of a special \emph{quantum circuit} applied to a simple fiducial state. This reverse description is not in violation with the irreversibility of the RG, since it only applies to the ground state, not to the whole Hilbert space and the Hamiltonian acting upon it. At stage $0$ of the MERA construction, the degrees of freedom (distinguishable quantum spins in this case) are arranged in a regular lattice and initialised in some convenient initial state, e.g., the ``all $0$s'' state $\ket{\mathbf{0}}$. At stage $1$ the degrees of freedom are subjected to a \emph{local interaction} $U_1$ for some constant time, resulting in a correlated quantum state $U_1\ket{\mathbf{0}}$ of the lattice. The precise details of $U_1$---while playing a key role in applications of the variational principle---are not required to establish the general properties of a MERA. At the next stage the lattice is subjected to a \emph{scale transformation} and the lattice spacing is doubled.  At this point, in order to \emph{renormalize} the lattice (i.e., restore the lattice spacing), new degrees of freedom are introduced: one adds a quantum spin initialised in the state $\ket{0}$ between each pair of the old lattice sites. We write this \emph{renormalization} step as $\mathcal{R}$. The resulting state $\mathcal{R}U_1\ket{\mathbf{0}}$ is then again subjected to a local interaction $U_2$ for some constant time followed by the renormalization step $\mathcal{R}$, producing the state $\mathcal{R}U_2\mathcal{R}U_1\ket{\mathbf{0}}$. This process is then iterated as many times as desired/required. A physical interpretation of the MERA construction is straightforward. Every layer $k$ begins with a dilation of the state living in the previous layer to a lattice doubled in size, by applying $\mathcal{R}$. The operation of $U_{k}$ then adds short-range (\textit{e.g.}\ over two sites) fluctuations/entanglement on top of this state, corresponding to fluctuations over $2^{m-k}$ sites in the final state if $m$ layers are applied. However, the resulting state $\ket{\Psi_{\text{MERA}}} = U_m \mathcal{R} U_{m-1} \mathcal{R} \cdots \mathcal{R} U_1 \ket{\mathbf{0}}$ generally breaks translation invariance unless the $U_j$ are carefully fine tuned.



\paragraph{The passage to the continuum.---}\hspace{-1em} For simplicity we specialise to the case of a single bosonic species in one dimension. (The generalisation to higher dimensions and to fermions or Bose-Fermi mixtures is entirely straightforward.) We write $\psi(x)$ and $\psi^\dag(x)$, $x\in \mathbb{R}$, for the bare field annihilation and creation operators which obey the canonical commutation relations $[\psi(x), \psi^\dag(y)] = \delta(x-y)$. The following constructions can also be described in terms of any set of operators that define the theory, such as the hermitian field operator $\phi$ and its conjugate momentum $\pi$ for relativistic boson theories.

The generalisation of the $0$th stage of the MERA construction is clear: one should choose for the initial state a factorized reference state $\ket{\Omega}$. The transition to a continuous space $x\in \mathbb{R}$ enables the introduction of a continuous scale parameter $s$ that labels the layers of the MERA construction. In every layer, new fluctuations are created by the action of a unitary evolution $U(s)=\exp(-\ic \delta K(s))$ with time step $\delta$ and \emph{local} interaction $K(s)$ given by
\begin{equation}
	K(s) = \int  k(x,s) \, dx.
\end{equation}
where $k(x,s)$ is a \emph{local} combination of the field operators $\psi(x)$, their adjoints, and their derivatives. Since the continuum lacks a shortest distance, we need to introduce a characteristic lengthscale $\epsilon$ below which $K(s)$ does not create entanglement. This is possible in a variety of ways: one strategy is to build $k(x)$ from \emph{smoothed} field operators $\widetilde{\psi}_\epsilon(x) = \int \chi_\epsilon(x-y)\psi(y)\,dy$, where $\chi_\epsilon(x)$ is some smooth envelope function which is nonzero outside a region of width $\epsilon$ around $x=0$. Another strategy is to simply impose a cutoff on $K$ when it is expressed in terms of momentum variables. The cutoff $\Lambda\approx \epsilon^{-1}$ ensures that the only degrees of freedom which are nontrivially affected are those with momenta $k\lesssim \Lambda$.
 
The third ingredient of the MERA construction is the \emph{renormalization step} where the scale is changed and new uncorrelated degrees of freedom are introduced. Here the continuum analogue is not entirely clear, but we argue that the following replacement naturally achieves the same objective: we simply effect a small change of scale via $W=\exp(-\ic \delta L)$, where the generator $L$ is given by 
\begin{equation}
	L = -\frac{i}{2}\int  \psi^{\dag}(x) x \frac{d\psi(x)}{dx} -   x\frac{d\psi^\dag(x)}{dx}\psi(x) \, dx.\label{eq:scalingoperator}
\end{equation}
This has the same physical outcome as the original MERA renormalization step because initially uncorrelated degrees of freedom at lengthscales below $\epsilon$ are now introduced at the new lengthscale. Thus our proposal for the continuous MERA (cMERA) is as follows: evolve some initial state $\ket{\Omega}$ according to $K$ for an infinitesimal time $\delta$, which correlates real-space degrees of freedom at scales of $\mathcal{O}(\epsilon)$. Then introduce new degrees of freedom from the higher momenta/shorter lengthscales by dilating the state via evolution according to $L$ for a time $\delta$. The last layer $s=s_{\epsilon}$ creates fluctuations at lengthscale $\epsilon$. No fluctuations at a shorter range exist in the final state. The fluctuations created by layer $s$ live at lengthscale $\epsilon \exp(s_{\epsilon}-s)$ in the final state. If this process is to correctly produce the long-range entanglement in a state with correlation length $\xi$ the first layer should be defined at $s=s_{\xi}$ with $s_{\epsilon}-s_{\xi}=\mathcal{O}(\log(\xi/\epsilon) )$. By taking the limit $\delta\rightarrow0$ we obtain our final expression:
\begin{equation}\label{eq:cMERA}
	\ket{\Psi} \equiv \mathcal{T}e^{-i\int_{s_\xi}^{s_\epsilon} K(s) + L \, ds}\ket{\Omega} \equiv U(s_{\epsilon},s_{\xi})\ket{\Omega}
\end{equation}
where $\mathcal{T}$ is the time-ordering operation. We refer to the unitary operation preparing a cMERA as $U(s_\epsilon, s_\xi)$. Note that a UV cutoff $\Lambda=\epsilon^{-1}$ and an IR cutoff $\xi^{-1}$ are explicitly built into the cMERA definition. For critical systems or relativistic theories with an infinite range of fluctuations it is necessary for $s_{\epsilon}-s_{\xi}\to \infty$.

The biggest difference between the cMERA and MERA definitions arises from the flexibility we have in imposing the UV cutoff. In the MERA case the UV cutoff is dictated by the lattice spacing. This, in turn, essentially forces the form of the subsequent scaling transformation (i.e., an integral number of spins must be introduced in the scaling step). The freedom, arising from the continuum, to choose a smooth UV cutoff allows the scaling transformation to be applied continuously and for the resulting state to be easily chosen to be translation invariant. 

The set of all cMERA forms a variational class: the variational parameters are the coefficients of interactions involved in $K$; these coefficients may depend on both $x$ and the integration parameter $s$. When the coefficients do not depend on position $x$ a generic cMERA is manifestly translation invariant (this is  established by noting that application of the unitary $e^{i\delta L}$ to a translation-invariant state results in a translation invariant state). This is in contradistinction to the generic situation with lattice MERA.

\paragraph{Comparison with Wilsonian renormalization.---}\hspace{-1em} The cMERA has been constructed using the quantum circuit interpretation, but can now be interpreted as an active renormalization process and compared to Wilson's momentum-shell renormalization group (RG) \cite{wilson}. The latter works by iteratively integrating out all the modes living in a small momentum shell $\Lambda - \d \Lambda < |\vk| < \Lambda$. A rescaling step brings modes at $\Lambda-\d \Lambda$ back to $\Lambda$. In the end $\Lambda$ can be sent to infinity, but we need to start with a finite $\Lambda$ in order to define the process. The renormalization process defined by the cMERA proposal is a \emph{real-space} implementation of Wilson's momentum shell RG in a \emph{hamiltonian} framework: rather than integrating out high-frequency modes around the cutoff $\Lambda$ from the partition function ---which is a fixed operation--- the operator $K$ first \emph{disentangles} these modes from the wavefunction in such a way that they can be isometrically projected onto a reference vacuum $\ket{\Omega}$. Then a scale transformation is performed to send the disentangled modes beyond the cutoff (where they no longer interact via $K$) and new entangled modes are brought to $\Lambda$. These modes are then disentangled in the subsequent step. This immediately clarifies the need for a finite cutoff $\Lambda$ in our construction (which can also be sent to infinity at the end of the process).

So what are the differences between the renormalization process defined by the cMERA and Wilson's momentum-shell RG? Firstly, whereas Wilson's RG is a fixed operation, the cMERA renormalization process is governed by $K(s)$ which can be variationally optimized. Secondly, while $K(s)$ can be formulated in momentum space (\textit{e.g.} to implement the cutoff), its defining property is real-space locality, which has proven to be a correct assumption for both free and interacting theories in countless examples with MERA and related variational classes and is a result of the locality of physical interactions.

\paragraph{The cMERA RG flow.---}\hspace{-1em} New to our formalism is that we can define the RG flow for operators in a hamiltonian framework: suppose we want to evaluate the expectation value $\braket{ O} = \braket{ \Psi| O |\Psi}$, where $\ket{\Psi}$ is a cMERA of the form Eq.~(\ref{eq:cMERA}), and $O$ is some local operator, e.g. $O = \psi(0)$ or $O = \psi^\dag(0) \psi(x)$.
To do this we define $O(s) \equiv U(s_\epsilon, s)^\dag O U(s_\epsilon, s)$, where
\begin{equation}\label{eq:aheispic}
	\frac{dO(s)}{ds} = -i[K(s) + L, O(s)].
\end{equation}
This `heisenberg-like' equation of motion is obtained by differentiating the \emph{lower limit} of the evolution operator $U(s_{\epsilon},s)$. Then $\braket{ O}$ may be found by integrating this equation from $s=s_\epsilon$ \emph{down} to $s = s_\xi$, with the initial condition $O(s_\epsilon) = O$, and evaluating $\braket{ \Omega|O(s_\xi)|\Omega}$.

Physically, we think of the \emph{bare} or \emph{physical} operator $O$ as being defined at the UV cutoff lengthscale $x \sim \epsilon$. As ``time'' $s$ progresses we think of $O(s)$ as being brought from lengthscale $x\sim \epsilon$ to lengthscale $x\sim \epsilon e^{s_{\epsilon}-s}$. Thus, $O(s)$ is obtained from the the bare operator $O$ by renormalizing up to scale $s$ (\textit{i.e.}\ all degrees of freedom between momentum scales $e^{s-s_{\epsilon}}\Lambda$ and $\Lambda$, where $s<s_{\epsilon}$, have been integrated out/disentangled). 

For critical theories, $K(s)$ is expected to become $s$-independent away from $s_{\epsilon}$. In accordance with the results from \cite{merascaling}, we can then assume the existence of operators $O$ that satisfy $-\ic[K+L,O]=\lambda O$. These are scaling operators with scaling dimension $\lambda$. If $O$ is a local scaling operator, it is necessarily centered around $x=0$. A local scaling operator $O(x)$ with scaling dimension $\lambda$ at position $x$ satisfies
\begin{equation}
-\ic[K+L,O]=x d O(x) / dx + \lambda O(x).\label{eq:scaling}
\end{equation}
The existence of scaling operators makes it easy to prove that cMERA support algebraically decaying correlations and are thus well suited to study critical theories. In addition, we can illustrate that they are able to produce an area law for the entanglement entropy. 

\paragraph{An entropy/area law for cMERA.---}\hspace{-1em} We now provide a heuristic argument that a generic cMERA obeys an entropy/area law by appealing to results \cite{entgen} concerning the dynamics of quantum spin systems: it is known that the \emph{entanglement entropy} $S_A(t) = -\tr(\rho_A(t)\log(\rho_A(t))$, where $\rho_A(t) = \tr_{A^c}(e^{-itH}\ket{\phi_0}\bra{ \phi_0}e^{itH})$ is the reduced density operator for a region $A$, under the real-time dynamics generated by \emph{any} strongly interacting system grows as 
\begin{equation}\label{eq:entropygrowth}
dS_A(t)/dt \le c|\partial A|, 
\end{equation}
where $|\partial A|$ denotes the length or area of the boundary of $A$ and should be measured in terms of the cutoff of $H$, for some constant $c$ which depends only on the local interactions and the geometry of $A$. It is natural to conjecture that this result holds in the continuum setting. Now, subject to this assumption, we can  deduce the proposed area law: we track the entropy growth of the time-dependent region $A(s) = A e^{s-s_\epsilon}$, i.e.\ $A(s)$ is $A$ scaled down by a factor $e^{s-s_\epsilon}$ throughout the cMERA preparation. We bound the entropy $S_A$ by integrating Eq.~(\ref{eq:entropygrowth}) (compare with \cite{vidalentanglemententropy}):
\begin{equation}
	\begin{split}
	S_A &\le c\int^{s_{\epsilon}}_{s_{\epsilon}-\log(L\Lambda )} (L\Lambda e^{s-s_{\epsilon}})^{d-1} \, ds\\ &= \begin{cases} c\log(L\Lambda), &\quad d=1 \\ \frac{c}{d-1}(L\Lambda)^{d-1}\left(1- \frac{1}{(L\Lambda)^{d-1}}\right), &\quad d>1, \end{cases} 
	\end{split}
\end{equation}
where the area $|\partial A|\leq(L\Lambda)^{d-1}$; the entanglement of $A(s)$ with the remainder of the system cannot receive further contributions when $|A(s)| \leq (L\Lambda)^{d}< 1$. The appearance of the logarithm of the UV cutoff is familiar from standard QFT calculations \cite{blackhole,cft}. Note that the cMERA might not describe logarithmic violations of the boundary law in $d>1$, similar to the MERA case.
  
\paragraph{Representation of ground states via cMERA.---}\hspace{-1em}
Let us now construct a cMERA representation of the ground state of a simple non-relativistic bosonic model 
\begin{displaymath}
\textstyle
H = \int \left[\frac{d\psi^{\dag}}{dx}\frac{d\psi}{dx} + \mu \psi^\dag\psi - \nu ({\psi^\dag}^2 + \psi^2)\right]\,dx.
\end{displaymath}
which is well-defined if $2\nu \leq \mu$. For $2\nu=\mu$, the elementary excitation becomes massless and the model becomes critical. Using a general strategy discussed in \cite{inpreparation}, we can describe ground states of free theories with a Gaussian cMERA where $K(s)$ is the quadratic operator
\begin{displaymath}
\textstyle
K(s)= -\frac{i}{2}\int g(\frac{k}{\Lambda},s)\left[\widehat{\psi}^\dag(k)\widehat{\psi}^\dag(-k)-\widehat{\psi}(-k)\widehat{\psi}(k)\right] dk,
\end{displaymath}
where $\widehat{\psi}(k)$ is the Fourier transform of the field operators. We set $g(k/\Lambda,s)=\chi(s) \Gamma(|k|/\Lambda)$ with $\Gamma(\kappa)$ a fixed cutoff function with cutoff $1$. The variational parameters are thus the function values $\chi(s)$. $\ket{\Omega}$ is fixed by $\psi\ket{\Omega}=0$. The analytic calculations are facilitated using a sharp cutoff such as $\Gamma(\kappa)=\theta(1-|\kappa|)$, where $\theta(x)$ is the Heaviside function. Although this cutoff function produces a nonlocal operator $K$, it is straightforward to see that similar results are obtained using a smooth cutoff such as $\Gamma(\kappa)=\exp(-\kappa^{2})$ which does yield a local $K$. The exact ground state of $H$ can be accurately reproduced if $\Lambda^{2}\gg \mathcal{O}(\mu)$ by choosing
\begin{displaymath}
\chi(s) = 2(\nu/\Lambda^{2}) / \left[ (e^{2s}+\Delta^{2}/\Lambda^{4} e^{-2s}) + 2\mu/\Lambda^{2}\right],
\end{displaymath}
where we have set $s_{\epsilon}=0$ and $\Delta=(\mu^{2}-4\nu^{2})^{1/2}$ represents the mass gap of the system. For $2\nu<\mu$ or thus $\Delta >0$, the `disentangling strength' $\chi(s)$ peaks at $s=-1/2\log(\Lambda^{2}/\Delta)$ and decays to zero for $s\to -\infty$. The integration of the RG flow can be stopped at $s_{\xi}\ll -1/2\log(\Lambda^{2}/\Delta)-\mathcal{O}(\log(\mu/\Lambda^{2}))$. In the critical limit ($2\nu \to \mu$), $\chi(s)$ reaches a non-zero horizontal asymptote $\lim_{s\to-\infty}\chi(s)=\nu/\mu=1/2$ and we need to integrate all the way down to $s_{\xi}=-\infty$. According to Eq.~\eqref{eq:scaling}, the scaling operators then correspond to $\phi(x)\sim (\psi(x)+\psi^{\dagger}(x))$ and $\pi(x)\sim(\psi(x)-\psi^{\dagger}(x))$. The low-energy behavior is scale-invariant and can be described by the massless Klein-Gordon equation \cite{inpreparation}.

\paragraph{Conclusions.---}\hspace{-1em}In this Letter a generalisation, cMERA, of the MERA variational class to the continuum setting has been introduced. We have argued that cMERA can be translation invariant and generically exhibit an entropy/area law. We have also supplied an analytic argument that the ground states of a general class of local quadratic models admits a cMERA description. Much remains to be done: we expect, by analogy with MERA, that cMERA will be a useful variational class for strongly interacting quantum fields, and will allow the description of a variety of interesting physical phenomena from topological effects to confinement, and symmetry breaking. Looking further afield, the cMERA constitutes a realization of the holographic principle. It is tempting to speculate, building on \cite{meraadscft} and \cite{tnsandgeom}, that cMERA are a natural candidate to establish a link between entanglement renormalization and the best known realization of the holographic principle, namely the AdS/CFT correspondence.

This work was supported by the Research Foundation Flanders (JH) and ERC grants QUERG and COQUIT. JH is grateful to TJO for the invitation to the Leibniz Universit\"at Hannover.

\clearpage

\begin{thebibliography}{99}
	\bibitem{wilson}
	K.~G.~Wilson, {Rev. Mod. Phys.} {\bf 47}, 773 (1975).
	
	\bibitem{vojta}
	M.~Vojta, N.-H.~Tong, and R.~Bulla, {Phys. Rev. Lett.} {\bf 94}, 070604 (2005).

	\bibitem{DMRG}
	S.~R.~White, {Phys. Rev. Lett.} {\bf 69}, 2863 (1992);
	U.~Schollw{\"o}ck, {Rev. Mod. Phys.} {\bf 77}, 259 (2005).
	
	\bibitem{MPS} 
	M.~Fannes, B.~Nachtergaele, R.~F.~Werner, {Commun. Math. Phys.} {\bf 144}, 443 (1992);
	S.~\"Ostlund  and S.~Rommer, {Phys. Rev. Lett.} {\bf 75}, 3537 (1995);
	F.~Verstraete, J.~I.~Cirac, V.~Murg, {Adv. Phys.} {\bf  57}, 143 (2008);
	J.~I.~Cirac, F.~Verstraete, {J. Phys. A: Math. Theor.} {\bf  42}, 504004 (2009).
	
	\bibitem{arealaws}
	F.~Verstraete J.~I.~Cirac, {Phys. Rev. B} {\bf 73}, 094423 (2006);
	T.~J.~Osborne, {Phys. Rev. Lett.}\ {\bf 97}, 157202 (2006);
	M.~B.~Hastings, {J. Stat. Mech.} P08024 (2007);
	J.~Eisert, M.~Cramer, M.~B.~Plenio, {Rev. Mod. Phys.} {\bf 82}, 277 (2010).
		
	\bibitem{tensornetworkrg}
	Z.-C.~Gu, M.~Levin, and X.-G.~Wen, {Phys. Rev. B} {\bf 78}, 205116 (2008);
	H.~H.~Zhao, Z.~Y.~Xie, Q.~N.~Chen, Z.~C.~Wei, J.~W.~Cai, and T.~Xiang, {Phys. Rev. B} {\bf 81}, 174411 (2010).
	
	\bibitem{ER}
	G.~Vidal, {Phys. Rev. Lett.} {\bf 99}, 220405 (2007);

	\bibitem{merascaling}
	V.~Giovannetti, S.~Montangero, R.~Fazio, {Phys. Rev. Lett.} {\bf 101}, 180503 (2008); R.N.C.~Pfeifer, G.~Evenbly and G.~Vidal, {Phys. Rev. A} {\bf 79}, 040301 (2009).
	
	\bibitem{mera}
	G.~Vidal, {Phys. Rev. Lett.} {\bf 101}, 110501 (2008).
	
	\bibitem{merageneral} 
	G.~Evenbly and G.~Vidal, {Phys. Rev. B} {\bf 79}, 144108 (2009);
    V.~Giovannetti, S.~Montangero, M.~Rizzi, and R.~Fazio, {Phys. Rev. A} {\bf 79}, 52314 (2009); 
	G.~Vidal, arXiv:0912.1651 (2009).

	\bibitem{mera2d}
	L.~Cincio, J.~Dziarmaga, and M.~M.~Rams, {Phys. Rev. Lett.} {\bf 100}, 240603 (2008); 
	G.~Evenbly and G.~Vidal, {Phys. Rev. Lett.} {\bf 102}, 180406 (2009); 
	G.~Evenbly and G.~Vidal, {Phys. Rev. Lett.} {\bf 104}, 187203 (2010).
	
	\bibitem{merafermions}
	P.~Corboz, G.~Evenbly, F.~Verstraete, G.~Vidal, Phys. Rev. A {\bf 81}, 010303(R) (2010); 
	C.~Pineda, T.~Barthel, and J.~Eisert, {Phys. Rev. A} {\bf 81}, 50303 (2010). 
	
	\bibitem{Bloch}
	I.~Bloch, J.~Dalibard, and W.~Zwerger, {Rev. Mod. Phys} {\bf 80}, 885 (2008).
	
	\bibitem{Impurities}
	P.~W.~Anderson, {Phys. Rev.} {\bf 124} 41 (1961).
	
	\bibitem{cMPS} 
	F.~Verstraete, J.~I.~Cirac, {Phys. Rev. Lett.} {\bf 104}, 190405 (2010);
	T.~J.~Osborne, J.~Eisert and F.~Verstraete, {Phys. Rev. Lett.} {\bf 105}, 260401 (2010).	
	
	\bibitem{RelcMPS} 
	J.~Haegeman, J.~I.~Cirac, T.~J.~Osborne, H.~Verschelde, F.~Verstraete, {Phys. Rev. Lett.} {\bf 105}, 251601 (2010); {PoS} {\bf FacesQCD}  029 (2010).
	
	\bibitem{fermiondoubling}
	H.~B.~Nielsen and M.~Ninomiya, {Nucl. Phys. B}{\bf 185}, 20 (1981);
	H.~B.~Nielsen and M.~Ninomiya, {Phys. Lett. B}{\bf 105}, 219 (1981).

	\bibitem{blackhole}
	L.~Bombelli, R.~K.~Koul, J.~Lee, and R.~D.~Sorkin, {Phys. Rev. D} {\bf 34}, 373 (1986);
	M.~Srednicki, {Phys. Rev. Lett.} {\bf 71}, 666 (1993).
	
	\bibitem{cft}
	C.~G.~Callan and F.~Wilczek, {Phys. Lett. B} {\bf 333}, 55 (1994);
	C.~Holzhey, F.~Larsen, and F.~Wilczek, {Nucl. Phys. B} {\bf 300}, 377 (1988);
	P.~Calabrese and J.~Cardy, {J. Stat. Mech.} P06002 (2004). 

	\bibitem{entgen}
	S.~Bravyi, M.~B.~Hastings, and F.~Verstraete, {Phys. Rev. Lett.} {\bf 97}, 050401 (2006);
	J.~Eisert, T.~J.~Osborne, {Phys. Rev. Lett.} {\bf 97}, 150404 (2006).
	
	\bibitem{vidalentanglemententropy}
	G.~Vidal, arXiv:quant-ph/0610099
	
	
	\bibitem{inpreparation}
	In preparation.

	\bibitem{meraadscft} 
	B.~Swingle, arXiv:0905.1317 (2009).
	
	\bibitem{tnsandgeom}
	G.~Evenbly and G.~Vidal, arXiv:1106.1082 (2011).
	

\end{thebibliography}
\end{document}